\documentclass[amsmath,amssymb,floatfix,prl]{revtex4}

\usepackage{graphicx}
\begin{document}

\title{First-order phase transition by a  spin-flip potential in BCS superconductivity}

\author{Octavio D. Rodriguez Salmon \footnote{Corresponding author.\\ E-mail address: octaviors@gmail.com}}
\affiliation{Universidade Federal do Amazonas, Departamento de F\'{\i}sica, 3000, Japiim, 69077-000, Manaus-AM, Brazil}

\author{Igor Tavares Padilha }
\affiliation{Universidade Federal do Amazonas, Departamento de F\'{\i}sica, 3000, Japiim, 69077-000, Manaus-AM, Brazil}

\author{Francisco Din\'{o}la Neto } 
\affiliation{Centro Universit\'{a}rio do Norte (UniNorte), Escola de Ci\^{e}ncias Exatas e Tecnologia, Manaus-AM, Brazil}

\begin{abstract}
A spin-flip potential $V$ is introduced  in a BCS Hamiltonian in order  to study a possible thermodynamic effect that might be caused by the  addition of a magnetic impurity in a superconductor. We found that the competition between the pairing and the spin-flip effect,  caused  by the interaction between the itinerant electrons and the spin of the impurity, gives rise to a first-order phase transition for certain values of $V$. The phase diagram  shows a tricritical point in a frontier line which is  topologically similar to those found in the phase diagrams of some Ising-like  models.

\textbf{PACS numbers}: 64.60.Ak; 64.60.Fr; 68.35.Rh
\end{abstract}

\maketitle

\section{Introduction}

The main effect of  magnetic impurities  in superconductors is the reduction of the critical temperature $T_{c}$, because the antiferromagnetic coupling between the magnetic moment in the impurity with the conduction electrons locally affects the time-reversal symmetry of the coherent pairing \cite{iavarone}. In other words, as the exchange interaction between the magnetic impurity and an electron is not time-reversal invariant, this kind of impurities oppose the pairing formation that mantains the superconductivity. Abrikosov and Gor'kov were the first ones who predicted theoretically the decreasing of $T_{c}$ if the impurities were only  magnetic \cite{abrikosov}, though it does not mean that  non-magnetic impurities can not  locally destroy the superconductivity \cite{li}. \\\\

The interaction $J\bf{S.s}$ between the spin $S$ of a localized impurity and the spin $s$ of an itinerant electron can be written in second-quantized form \cite{kasuya,jensen, hewson}  as follows:

\begin{equation}
{\cal H}_{e-imp} = \sum_{\bf{k,k^{\prime}}} J_{\bf{k,k^{\prime}}} \left \{ S^{\dagger}C_{{\bf k},\downarrow}^{\dagger}C_{{\bf k}^{'},\uparrow} +S^{-}C_{{\bf k},\uparrow}^{\dagger}C_{{\bf k}^{'},\downarrow}  + S_{z} (C_{{\bf k},\uparrow}^{\dagger}C_{{\bf k}^{'},\uparrow} - C_{{\bf k},\downarrow}^{\dagger}C_{{\bf k}^{'},\downarrow}  )\right \}
\label{imp}
\end{equation}
where $C_{{\bf k},s}^{\dagger}$ and $C_{{\bf k},s}$ are creation and anihilation operators for a conduction electron with spin $s$ in the ${\bf k}$-space, obeying anticonmutation relations. In this work we simplify this interaction in such a way that we only want to examine the effect of the spin flipping (see the first two terms in Eq.(\ref{imp})).  Furthermore, for the sake of simplicity, we consider only  elastic scatterings, so $k = k^{\prime}$. Thus, the Hamiltonian describing the pairing process affected by an spin-flip potential can be written as follows:

\begin{equation}
{\cal H} = \sum_{s,{\bf k}}E_{{\bf k}} C_{{\bf k},s}^{\dagger}C_{{\bf k},s} - U \sum_{{\bf k},{\bf k}^{'}}C_{{\bf k}^{'},\uparrow}^{\dagger}C_{-{\bf k}^{'},\downarrow}^{\dagger}C_{-{\bf k},\downarrow}C_{{\bf k},\uparrow} + \sum_{s,{\bf k}} V_{{\bf k}} C_{{\bf k},s}^{\dagger}C_{{\bf k},-s},
\label{hamiltoniano1}
\end{equation}
 The first sum stands for the energy of  free electrons, the second sum shows the attractive energy $U$ that generates the electron pairs with opposite spins and momenta, and the third one stands for the spin-flip mechanism due to the scattering potential $V_{{\bf k}}$ coming from an ad hoc simplification of the $J\bf{S.s}$ interaction.\\\\

 The suitable order parameter for measuring the number of electron pairs is defined  as: 

\begin{equation}
\displaystyle \Delta = U \sum_{{\bf k}} \langle C_{-{\bf k},\downarrow}C_{{\bf k},\uparrow} \rangle,
\label{parametrodeordem}
\end{equation}

where $\langle...\rangle$ is the thermal mean in the canonical ensemble. In order to study the effect of $V_{{\bf k}}$ on $\Delta$, we firstly need to calculate this thermal mean by using
the Zubarev's Green Function technique \cite{zubarev}, in which $\langle C_{-{\bf k},\downarrow}C_{{\bf k},\uparrow} \rangle$ is obtained through the following integral:

\begin{equation}
\displaystyle \langle C_{-{\bf k},\downarrow}C_{{\bf k},\uparrow} \rangle = 2 \int d\omega  [ \frac{ {\bf Im} \tilde{G}(\omega)}{\exp(\beta \omega)+1} ], 
\label{salto}
\end{equation}

where  $\tilde{G}(\omega)$ is the Fourier transform in the $\omega$-space of 

\begin{equation}
\displaystyle G(t-t^{\prime}) = \langle \langle C_{-{\bf k},\downarrow} (t) ; C_{{\bf k},\uparrow}(t^{\prime}) \rangle \rangle =\frac{1}{i}\theta(t-t^{\prime}) \langle \{  C_{-{\bf k},\downarrow} (t) , C_{{\bf k},\uparrow}(t^{\prime}) \} \rangle , 
\end{equation}

which is the retarded Green Function that has to be determined so as to calculate the integral in Eq.(\ref{salto}). To this end, a first step is to find the evolution of $ G(t-t^{\prime})$ by deriving the above equation:

\begin{equation}
\frac{d}{dt}G(t-t^{\prime})  = \frac{1}{i}\delta(t-t^{\prime})
\langle \{C_{-{\bf k},\downarrow} (t),C_{{\bf k},\uparrow}(t^{\prime}) \} \rangle + \frac{1}{i}\theta(t-t^{\prime})\langle \{ \frac{d}{dt}C_{-{\bf k},\downarrow}(t), C_{{\bf k},\uparrow}(t^{\prime})\}\rangle.
\end{equation}

The first term is zero due to the delta function and a anticonmutator property. The derivative of the operator $C_{-{\bf k},-\sigma}(t)$ in the anticonmutator $\{...\}$ can be obtained by the Heisenberg equation:

\begin{equation}
 i \frac{d}{dt}C_{-{\bf k},\downarrow} = [C_{-{\bf k},\downarrow}(t), {\cal H}],
\end{equation}

so the equation of movement for the Green Function G is:

\begin{equation}
\frac{d}{dt}G(t-t^{\prime})  =  \frac{1}{i}\theta(t-t^{\prime})\langle \{ \frac{1}{i}[C_{-{\bf k},\downarrow}(t), {\cal H}] , C_{{\bf k},\uparrow}(t^{\prime})\}\rangle.
\label{movimiento}
\end{equation}

This generates an infinite number of equations unless we decouple  the term $C_{{\bf k}^{'},\uparrow}^{\dagger}C_{-{\bf k}^{'},\downarrow}^{\dagger}C_{-{\bf k},\downarrow}C_{{\bf k},\uparrow} $ in the Hamiltonian. 
The simplest way of approaching a product like  $ABCD$ is by considering only linear terms of fluctuations ($x- \langle x \rangle$):

\begin{equation}
ABCD = (\langle AB \rangle - (\langle AB \rangle -AB)) (\langle CD \rangle - (\langle CD \rangle -CD)) \simeq \langle AB \rangle CD + AB \langle CD \rangle - \langle AB \rangle \langle CD \rangle.
\end{equation}

Accordingly, by the use of the above approximation and considering Eq.(\ref{parametrodeordem}), the mean-field expression of the   Hamiltonian   given in Eq(\ref{hamiltoniano1}) is :

\begin{equation}
{\cal H}_{mf} = \sum_{s,{\bf k}}E_{{\bf k}} C_{{\bf k},s}^{\dagger}C_{{\bf k},s} - \Delta \sum_{{\bf k}} C_{{\bf k},\uparrow}^{\dagger}C_{-{\bf k},\downarrow}^{\dagger} 
- \Delta^{*}  \sum_{{\bf k}}C_{-{\bf k},\downarrow}C_{{\bf k},\uparrow} + \sum_{s,{\bf k}} V_{{\bf k}} C_{{\bf k},s}^{\dagger}C_{{\bf k},-s} + \frac{{|\Delta|}^{2}}{U}.
\end{equation}
This simpler Hamiltonian turns Eq.(\ref{movimiento}) as

\begin{equation}
\frac{d}{dt}G(t-t^{\prime})  = \varepsilon_{-{\bf k}} G(t-t^{\prime}) + \Delta_{{\bf k}} R(t-t^{\prime}) + V_{-{\bf k}} P(t-t^{\prime}),
\end{equation}

where

\begin{equation}
R(t-t^{\prime})=  \langle \langle C_{{\bf k},\uparrow}^{\dagger} (t) ; C_{{\bf k},\uparrow}(t^{\prime}) \rangle \rangle, 
\end{equation}
and 
\begin{equation}
P(t-t^{\prime})=  \langle \langle C_{-{\bf k},\uparrow} (t) ; C_{{\bf k},\uparrow}(t^{\prime}) \rangle \rangle
\end{equation}

In addition of these three Green Functions, the equation of movement of $R(t-t^{\prime})$ gives rise to another Green Function $Q(t-t^{\prime})$ given by:

\begin{equation}
Q(t-t^{\prime})=  \langle \langle C_{{\bf k},\downarrow}^{\dagger} (t) ; C_{{\bf k},\uparrow}(t^{\prime}) \rangle \rangle.
\end{equation}

So, the four generated movement equations for these Green Functions can be Fourier Transformed in the $\omega$-space, leading to four equations:

\begin{equation}
\omega \tilde{G}(\omega) = E_{-{\bf k}} \tilde{G}(\omega) + \Delta_{{\bf k}} \tilde{R}(\omega) + V_{-{\bf k}} \tilde{P}(\omega) ,
\end{equation}

\begin{equation}
\omega \tilde{R}(\omega) = \frac{1}{2\pi} - E_{{\bf k}} \tilde{R}(\omega) + {\Delta}_{-{\bf k}}^{*} \tilde{G}(\omega) - V_{{\bf k}} \tilde{Q}(\omega)
\end{equation}

\begin{equation}
\omega \tilde{P}(\omega)  = E_{-{\bf k}}\tilde{P}(\omega)  - \Delta_{{\bf k}} Q_{\omega} + V_{-{\bf k}} \tilde{G}(\omega)
\end{equation}

\begin{equation}
\omega \tilde{Q}(\omega) = -E_{{\bf k}}\tilde{Q}(\omega) -  {\Delta}_{-{\bf k}}^{*} \tilde{P}(\omega)  - V_{{\bf k}} \tilde{R}(\omega)
\end{equation}

We are considering here s-wave superconductivity, so $E_{{\bf k}} =  E_{-{\bf k}}$ and $\Delta_{{\bf k}}= \Delta_{-{\bf k}}=\Delta$. We assume the same symmetry for the spin-flip potential and ${\bf k}$ independency, thus $V_{{\bf k}} = V_{-{\bf k}} = V$. Accordingly, after solving the set of equations with these considerations, the expression of the Green Function of our  interest is conveniently written as follows:

\begin{equation}
\tilde{G}(\omega) = \frac{\Delta}{8\pi \sqrt{E_{{\bf k}}^{2}+\Delta^{2}}}\left ( \frac{1}{\omega-\omega_{1}} +\frac{1}{\omega-\omega_{2}} -\frac{1}{\omega+\omega_{1}} -\frac{1}{\omega+\omega_{2}} \right ),
\label{green}
\end{equation} 

where $\omega_{1} = V+\sqrt{E_{{\bf k}}^{2}+\Delta^{2}}$ and $\omega_{2} = -V+\sqrt{E_{{\bf k}}^{2}+\Delta^{2}}$. By substituting the expression of Eq.(\ref{green}) into Eq.(\ref{salto}), we have the integral equation:
\begin{equation}
\Delta = g(T,\Delta,V),
\label{gap1}
\end{equation}
where
\begin{equation}
g(T,\Delta,V) =  \frac{\rho(E_{F})U}{4} \Delta \int_{-\omega_{D}}^{\omega_{D}} \frac{\tanh(\frac{\beta \omega_{1}}{2})+\tanh(\frac{\beta \omega_{2}}{2})}{\sqrt{E^{2}+\Delta^{2}}}dE,
\label{defineg}
\end{equation}

being the Debye energy  $\omega_{D}$ in  $\hbar$ units and $\beta = 1/k_{B}T$, where we set $k_{B}=1$ for simplicity. Note that $\Delta$ can be out of the integral because we are considering s-wave pairing, so $\Delta$ does not depend on $\bf{k}$. In this case the factor  $\Delta$ can be canceled in the integral equation given in Eq.(\ref{gap1}) leading to the following gap equation:
\begin{equation}
1= f(T,\Delta,V),
\label{gap2} 
\end{equation}
where
\begin{equation}
f(T,\Delta,V) = g(T,\Delta,V)/\Delta  .
\end{equation}
This nonlinear equation can be solved  numerically so as  to obtain the gap $\Delta$, for given values of the temperature $T$ and the potential $V$.  However, for $T \neq 0$, only solutions that minimize the free energy must be acepted. Accordingly, it is necessary to calculate the free energy, which is obtained by integrating Eq.(\ref{gap1}) , so:

\begin{equation}
F(\Delta^{\prime},T,V) = F_{n}+ \frac{1}{2}{\Delta^{\prime}}^{2}- \int_{0}^{\Delta^{\prime}} g(T,x,V)dx, 
\label{fenergy}
\end{equation} 

where $F_{0}$ is the free energy in the normal state, i.e., for $\Delta =0$. Therefore, the gap equation gives us a thermodynamic stable solution of the $\Delta$ curve provided that the free energy $F$ has an absolute minimum for $\Delta^{\prime} = \Delta$, for given values of $T$ and $V$. If  that minimum is not an absolute one, we had  a spurious solution corresponding to a metastable state (see reference \cite{igor1}), so it would not belong to the equilibrium. This  analysis is  especially necessary if a first-order phase transition takes place, because Eq.(\ref{gap2}) can only works for  second-order points, i.e., points of continuity of $\Delta = \Delta(T,V)$. Thus,  if the curve $\Delta$ versus $T$ suffers a jump discontinuity to zero at some temperature $T_{c}$, for $\Delta^{\prime}=\Delta_{c}$ and for a given value of $V$,  the function $F$ versus $\Delta^{\prime}$ would exhibit two equal absolute minima, such that $F(T_{c},\Delta_{c},V) = F(T_{c},0,V)$. This signals a first-order phase transition where the superconducting phase ($\Delta \neq 0$) coexists with the normal phase ($\Delta = 0$). The same analysis has been done for other phase transition phenomena such as the magnetization \cite{octavio1}.    In the next section we study the phase diagram of this model applying this thermodynamic criterion.

\section{The Phase diagram}

The first step in obtaining  the phase diagram is to solve numerically the well known gap equation Eq.(\ref{gap2}) for $V=0$, considering that  $10^{-3} < \omega_{D} < 10^{-2}$ in the scale of the Fermi energy $E_{F}$. We set $\rho(E_{F})U = 0.6$, which agrees with  the experimental range of the BCS coupling.    In Figure 1 we observe the continuous behavior of the gap as a function of the temperature, indicating a second-order phase transition at a critical temperature $T=T_{c}$. We define $\Delta_{c}$ as the gap at zero temperature, which is approximately $1.77T_{c}$, thus we conveniently define the normalized  variables $\delta = \Delta/\Delta_{c}$ and $t=T/T_{c}$. \\\\
 The next step in exploring the phase diagram is to find the critical value of the spin-flip potential $V$ for which the critical temperature is zero. To this end, we take the limit $\beta \to \infty$  in Eq.(\ref{defineg}), so we have:

\begin{equation}
g(0,\Delta,V) = \frac{\rho(E_{F})U}{4} \Delta \int_{-\omega_{D}}^{\omega_{D}} \frac{1+sign(\sqrt{E^{2}+\Delta^{2}}-V)}{\sqrt{E^{2}+\Delta^{2}}}dE,
\end{equation}
Accordingly, the ground-state energy can be expressed as

\begin{equation}
E_{gs}= E_{n}+ \frac{1}{2}{\Delta^{\prime}}^{2}- \int_{0}^{\Delta^{\prime}} g(0,x,V)dx, 
\label{gsenergy}
\end{equation} 

where $E_{n}$ is the energy of the normal state at $T=0$, and $\Delta^{\prime}$ must be such that  $E_{gs}$ is minimized when $\Delta^{\prime}=\Delta$, so $\Delta$ is the value of the gap in equilibrium. Note that if we find that $E_{gs}$ has an absolute minimum at $\Delta =0$ (for a given values of $V$), we have $E_{gs}=E_{n}$, so it is convenient to plot the difference $E_{gs}-E_{n}$ versus $\Delta^{\prime}$, for given values of $V$, so as to determine for which value of $V$, say $V_{c}$, there is a phase transition between the superconducting and the normal state.  \\\\
In Figure 2 we may observe the difference $E_{gs}-E_{n}$ as a function of the normalized value of $\Delta^{\prime}$ ($\delta^{\prime} = \Delta^{\prime}/\Delta_{c}$), for three representative values of $V$. Thus,  in Figure 2a, we see that  the ground state is in the superconducting BCS state, because the energy reaches its minimum for a nonzero gap, for a chosen  value of $V$ that is a little bit less than a critical $V_{c}$ ($V = 0.957 V_{c}$). In Figure 2b is shown what happens with the energy when $V=V_{c}$ at $T=0$. The energy has two minima at the same level; one is for $\delta^{\prime} =0$ (normal state) and the other one is for $\delta^{\prime} =1$ (the BCS state). Therefore, at this point the system suffers a first-order quantum phase transition. We found that $V_{c}= 2.7155 \times 10^{-3}E_{F}$.  In Figure 2c, we may see that the energy has one minimum at $\delta^{\prime} = 0$, for a value of $V$ that is a little bit greater than $V_{c}$, signaling that the ground state is now in the normal state.   \\\\

Now, after knowing the phases of the ground state according to behavior of the gap, we can study the frontier that divides the BCS phase and the normal state for finite temperatures ($T > 0$). To this end we may observe the behavior of the gap, which is the order parameter, as a function of the temperature, for given values of $V$.  Conveniently, we use the normalized variable $\nu = V/V_{c}$, since we have seen that for $V > V_{c}$ the gap is zero, for $T=0$. Consequently, $\delta$ is zero, for $\nu > 1$, for $T>0$. So in Figure 3 we plotted three curves of the gap, for three representative values of $\nu$. Note that the vertical axis is represented by $\delta$ (not by $\delta^{\prime}$), because we are showing equilibrium values of the gap.  For $\nu = 0$, the gap curve is continuous, as also shown in Figure 1. For $\nu = 0.81$, the gap still falls continuously downto zero, signaling a second-order phase transition at a critical normalized  temperature $t=0.65$. These two curves were obtained by solving the gap equation given in Eq.(\ref{gap2}). However, this equation can not be used if the gap suffers a jump discontinuity when falling to zero. At this point a first-order phase transition would occur, so there would be a coexistence of the ordered phase with  the disordered phase. Thus, we must also  use the free energy given in Eq.(\ref{fenergy}), in order to determine the transition temperature for discontinuous change of phase.   \\\\

For $\nu = 0.884$, the gap curve exhibits a first-order phase transition, as shown in Figure 3. The transition temperature is approximately 0.44. We show it in Figure 4, where the difference $F-F_{n}$ (see Eq.(\ref{fenergy})) has been plotted as a function of $\delta^{\prime}$. There we may see that two minima are just at the same level, showing that there is a first-order phase transition approximately at $t = 0.44$, for that value of $\nu$. So we can infere that the phase diagram in the $\nu$-t plane must contain a frontier curve dividing the BCS and the normal state, with two sections separated by a tricritical point located at ($\nu^{*}$, $t^{*}$). Also, the first-order section of the  frontier curve must finish at $t=0$, for $\nu=1$, as analyzed before (see Figure 2). Then,  After observing Figure 3 we may infere that $ 0.810 < \nu^{*} < 0.884$. We estimated  $\nu^{*}$ approximately by scanning numerically the critical temperatures of the frontier, for different values of $\nu$. Accordingly, we found $(\nu^{*}, t^{*}) \simeq (0.86, 0.52)$. \\\\

Finally, we show in Figure 5 the phase diagram in the $\nu$-t plane, exhibiting a frontier line that divides the superconducting phase and the normal phase. The continuous line represents the second-order section of the frontier line, whereas the dotted line stands for the first-order one. The tricritical point is represented by the black circle. As may be noticed, the first-order line is perpendicular to the horizontal axis at $t=0$, which is in agreement with the second law of thermodynamics. 

\section{Conclusions}

We obtained numerically the phase diagram of a BCS Hamiltonian subjected to a spin-flip potential $V$. The analysis of the ground state showed that there is a critical value of $V$ (denoted by $V_{c}$), such that for $V< V_{c}$, the system is in the BCS phase, whereas for $V > V_{c}$ the normal state is present. For $V=V_{c}$ the system suffers a first-order phase transition due to the coexistence of two minima of the ground-state energy, at $\Delta = 0$ and $\Delta = \Delta_{c}$ ($\delta = 1$) (see Figure 3). Consequently, for finite temperatures, we found a frontier line having a tricritical point. The resulting curves were obtained carefully by ensuring the free energy minimization, which guarantees the thermodynamic equilibrium. Interestingly,  the phase diagram in the plane of temperature versus the spin-flip potential exhibits the same topology as found in phase diagrams of Ising-like Hamiltonians, such as the the Blume-Capel Model and the Random-Field Ising Model \cite{aharony, octavio2}. 
\\\\
As a perspective of this work it would be important  to consider the more realistic effect  of an inelastic scattering that flips the spin of the itinerant electron. This opens the question if the first-order line still remains for certain values of the spin-flip potential.

\vskip \baselineskip  
{\large\bf Acknowledgments}

Financial support from CNPq (Brazilian agency) is acknowledged. 

\vskip \baselineskip

\vskip \baselineskip
\vspace{3.0 cm}
\begin{figure}[htbp]
\centering
\includegraphics[height=6cm]{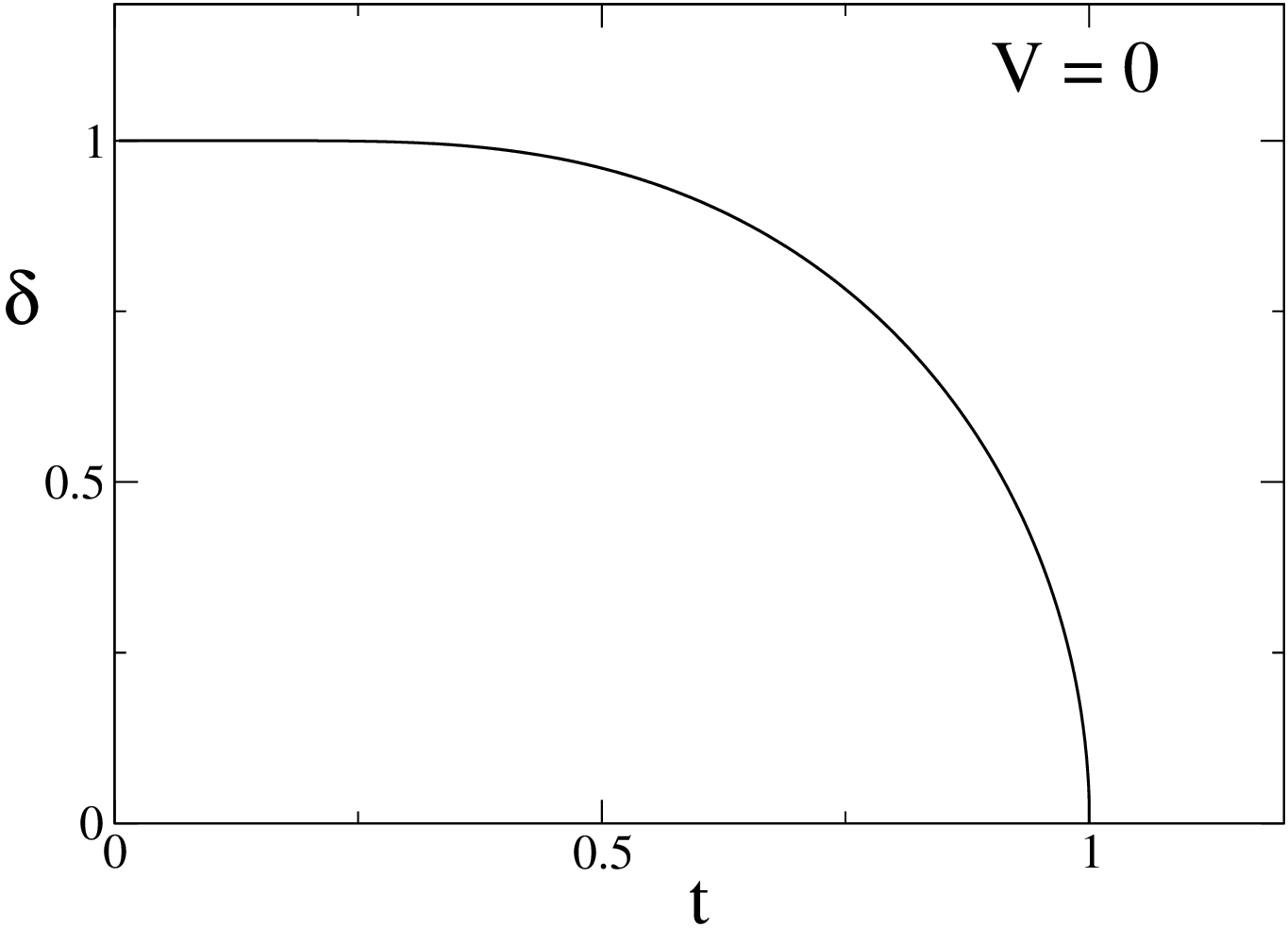}
\caption{The gap curve as a function of the temperature,  for $V =0$. The  axes have been normalized, where $\delta = \Delta/\Delta_{c}$ and $t = T/T_{c}$.  } 
\label{fig1}
\end{figure}

\vskip \baselineskip
\vspace{3.0 cm}
\begin{figure}[htbp]
\centering
\includegraphics[height=3.75cm]{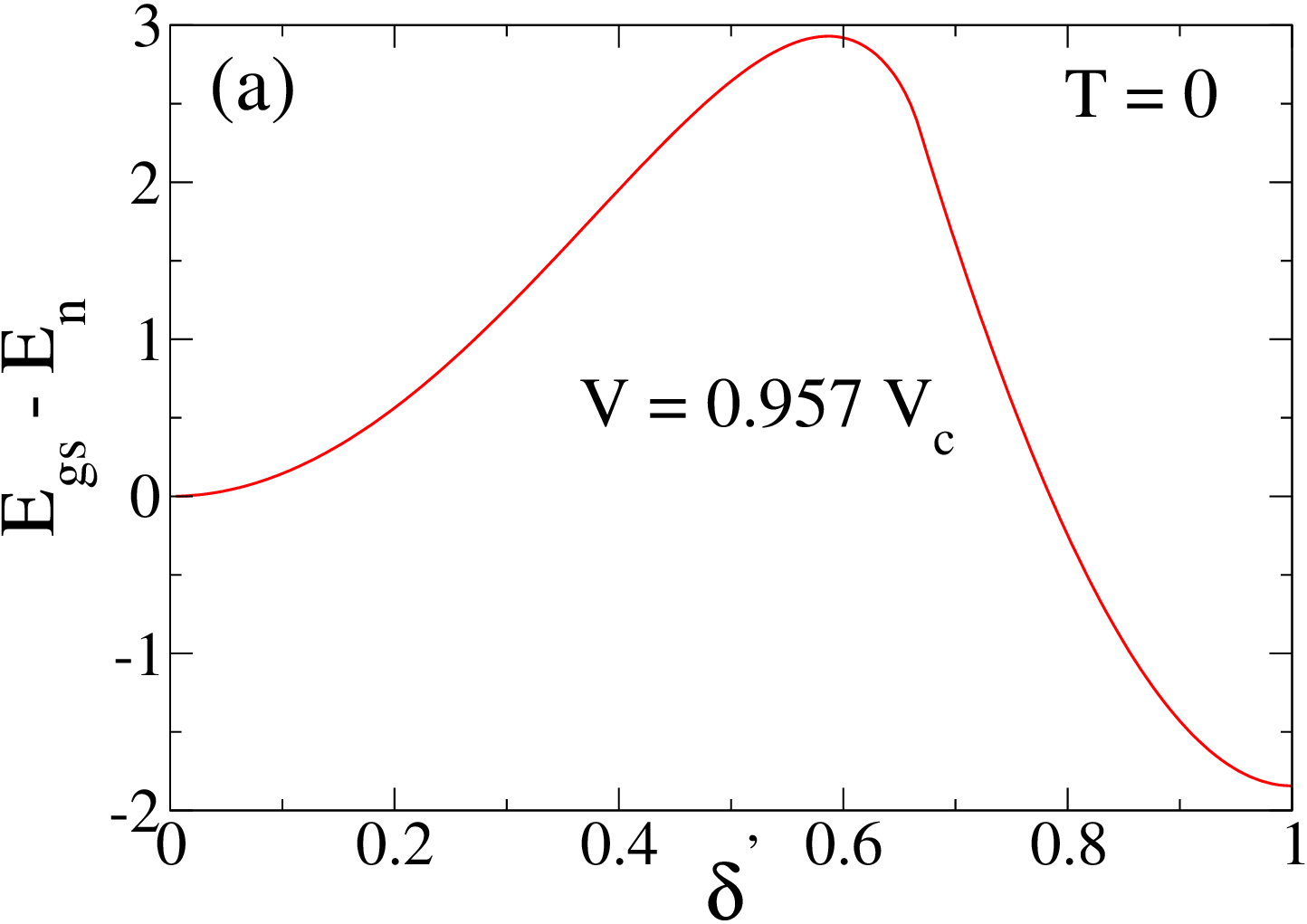}
\includegraphics[height=3.75cm]{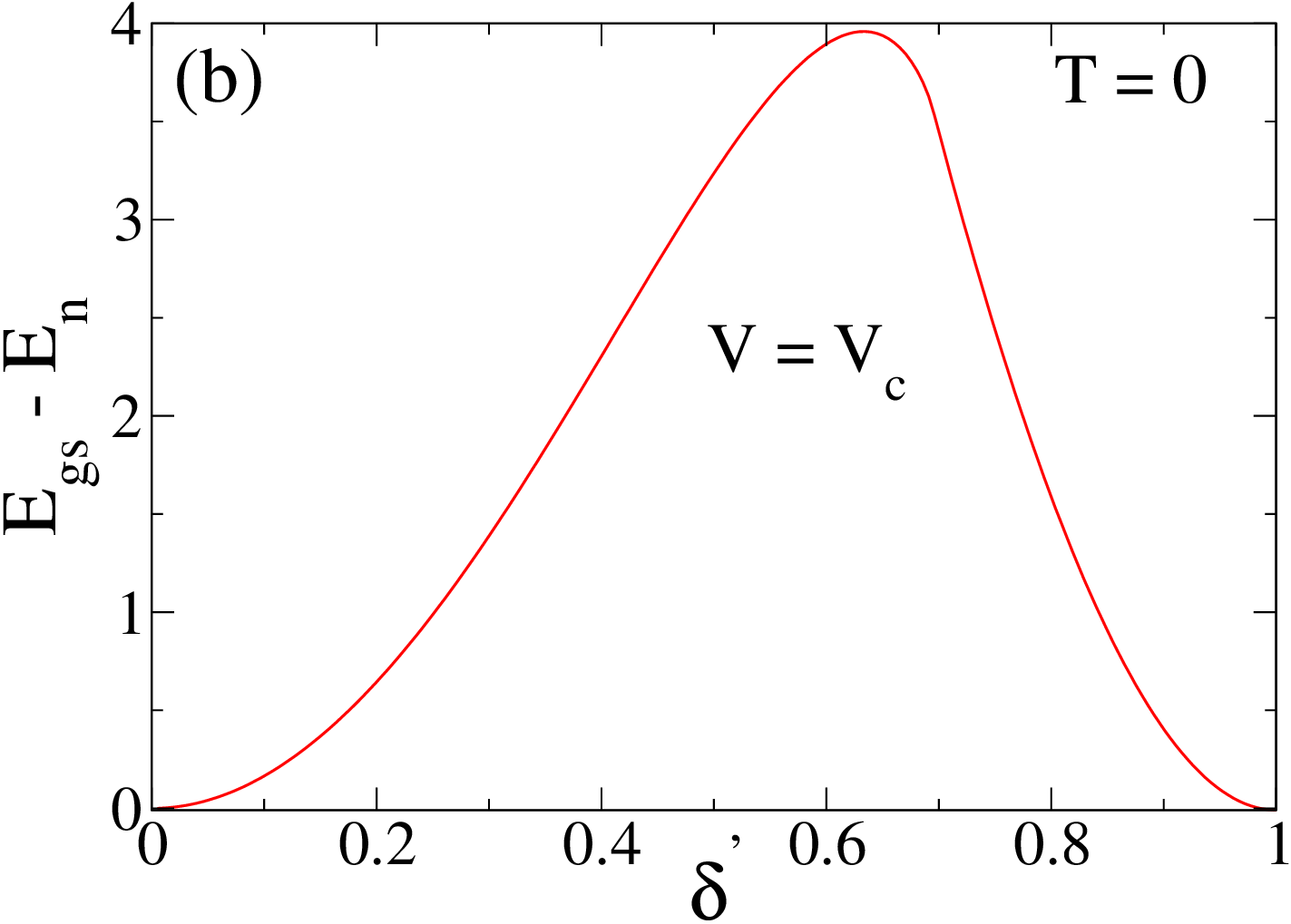}
\includegraphics[height=3.75cm]{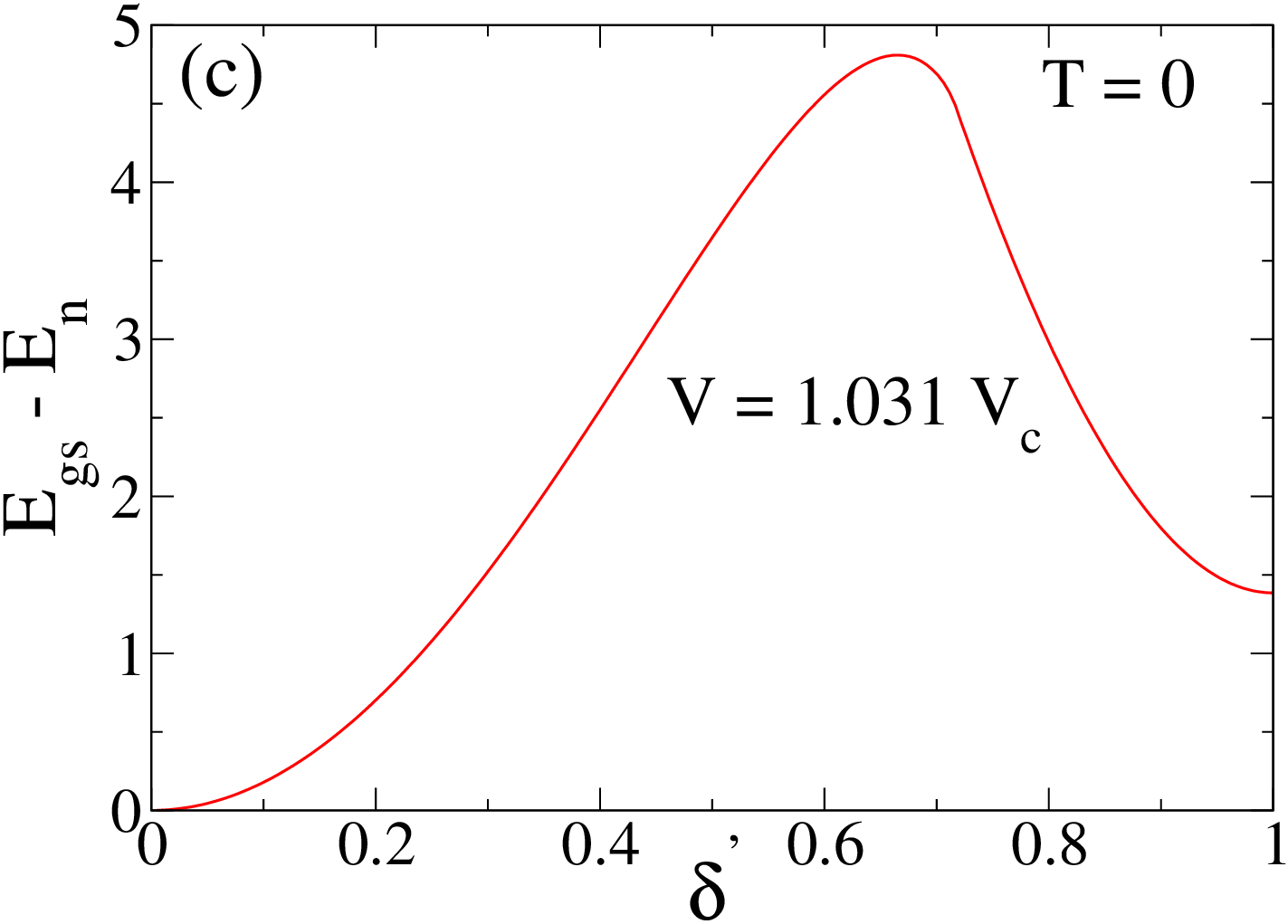}
\caption{The difference $E_{gs}-E_{n}$ (in convenient units) as a function of the normalized gap $\delta^{\prime}$, for three different values of $V$ at zero temperature. The gap that minimizes the energy determines the phase of the ground state. Accordingly,  the system is in the BCS state for $V < V_{c}$, and in the normal state for $V > V_{c}$. For $V=V_{c}$, the system suffers a first-order quantum phase transition. } 
\label{fig2}
\end{figure}

\vskip \baselineskip
\vspace{3.0 cm}
\begin{figure}[htbp]
\centering
\includegraphics[height=6cm]{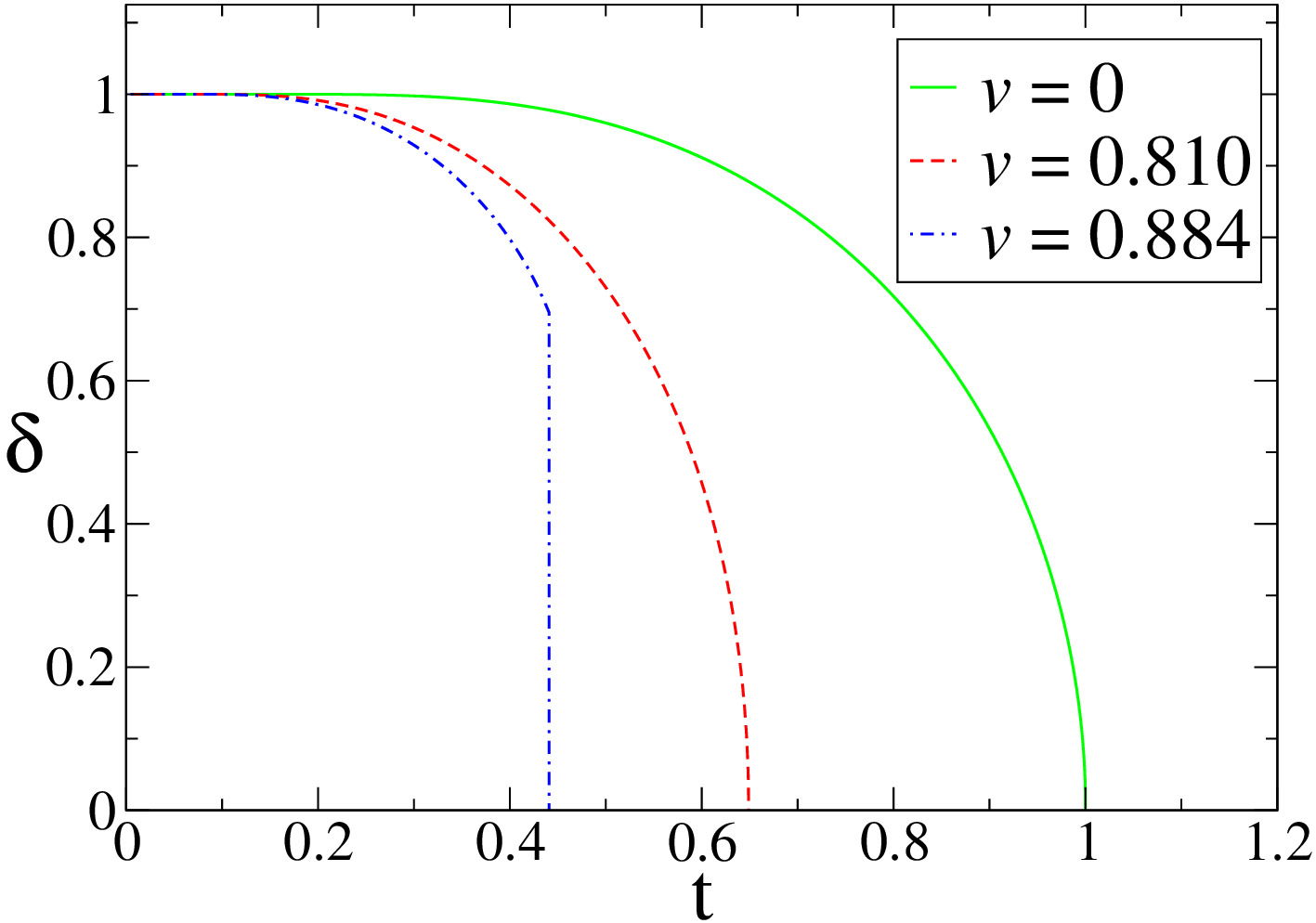}
\caption{The gap versus the normalized temperature, for three different values of the normalized potential $\nu$. It can be observed that 
the gap curve suffers a jump discontinuity at $t \simeq 0.44$, when $\nu = 0.884$. So there must be a critical value of $\nu$, for $0.810 < \nu < 0.884$,  for which we can locate the tricritical point dividing the second- and the first-order frontier of the phase diagram in the $\nu$-t plane. } 
\label{fig3}
\end{figure}

\vskip \baselineskip
\vspace{3.0 cm}
\begin{figure}[htbp]
\centering
\includegraphics[height=6cm]{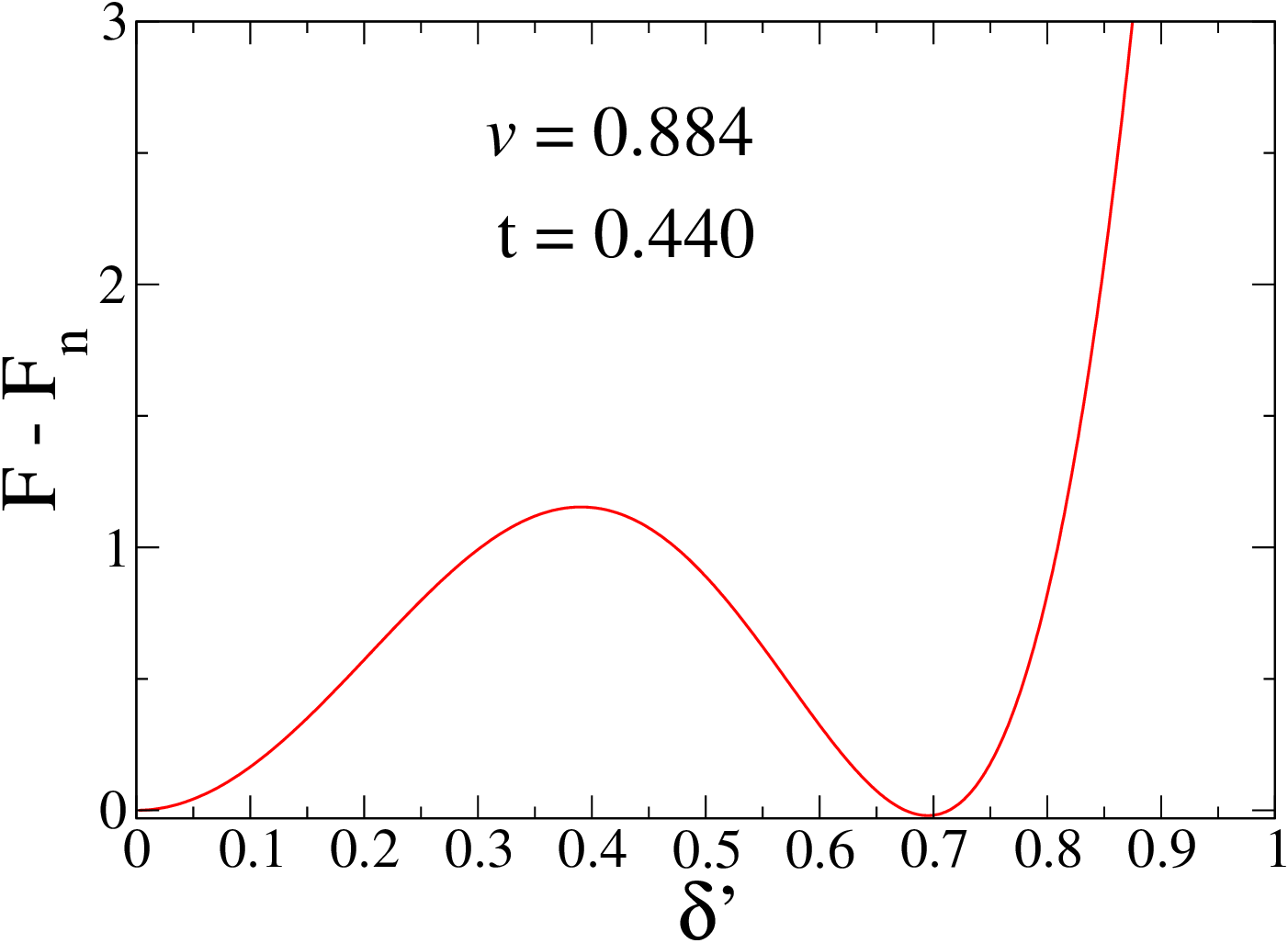}
\caption{The free energy minus the free energy of the normal state versus $\delta^{\prime}$, for $\nu = 0.884$ and $t=0.44$. Note that there are two minima just at the same level, for   $\delta^{\prime}=0$ and $\delta^{\prime} \simeq 0.7$. This means that the point ($\nu$,t) = (0.884,0.44) belongs to the first-order section of the frontier dividing the BCS and the normal state in the phase diagram in the $\nu$-t plane. The vertical axis is in convenient units. } 
\label{fig4}
\end{figure}

\vskip \baselineskip
\vspace{3.0 cm}
\begin{figure}[htbp]
\centering
\includegraphics[height=6cm]{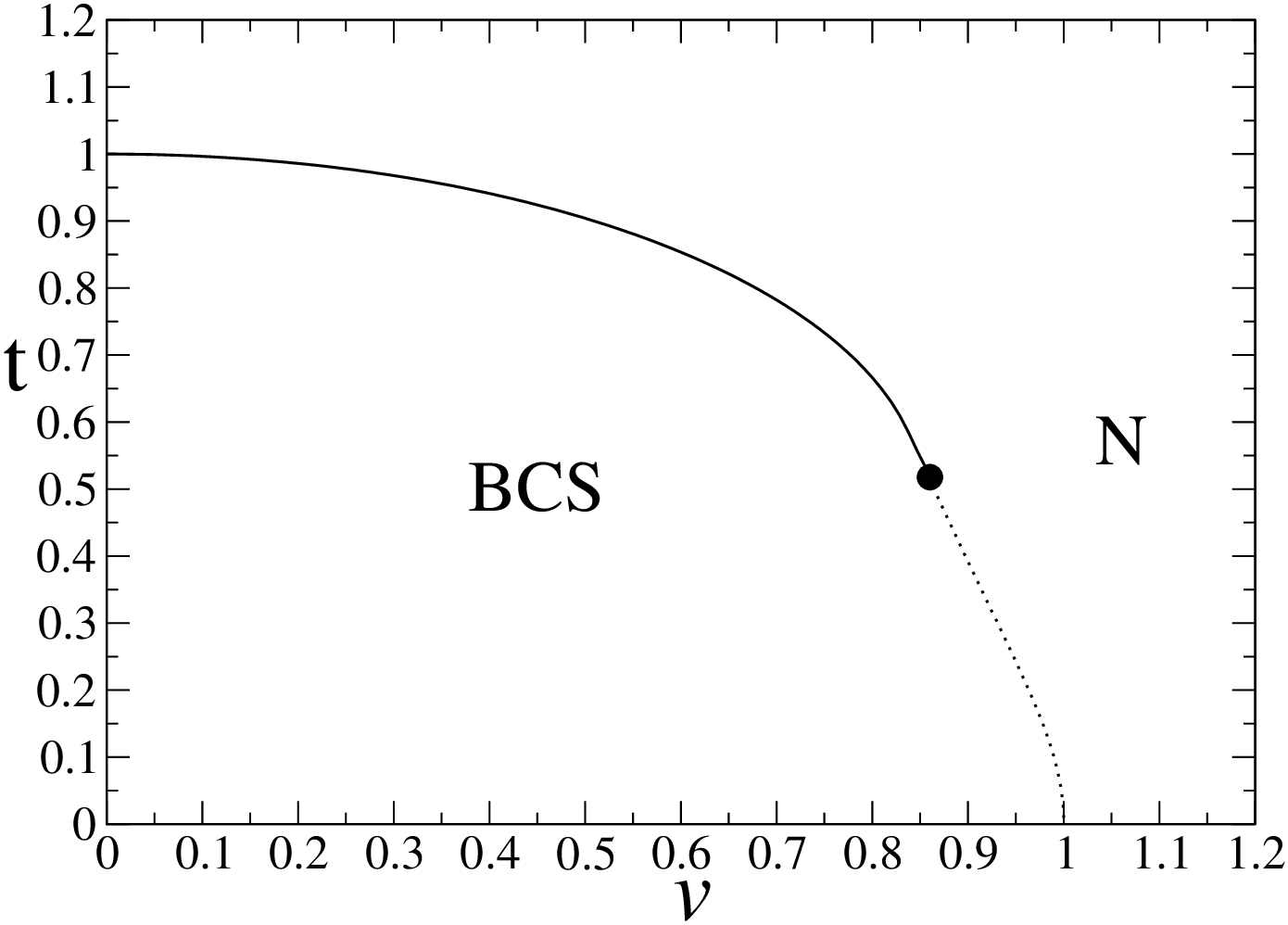}
\caption{Phase diagram in the $\nu$-t plane. The frontier dividing the BCS and the normal phases has two sections. The second-order and the first-order sections are represented by the continuous and the dotted line, respectively. The tricritical point is represented by the black circle, and is located at  $(\nu^{*}, t^{*}) \simeq (0.86, 0.52)$.  } 
\label{fig5}
\end{figure}

\end{document}